\magnification \magstep1
\raggedbottom
\openup 2\jot
\voffset6truemm
\def\cstok#1{\leavevmode\thinspace\hbox{\vrule\vtop{\vbox{\hrule\kern1pt
\hbox{\vphantom{\tt/}\thinspace{\tt#1}\thinspace}}
\kern1pt\hrule}\vrule}\thinspace}
\centerline {\bf A NEW FAMILY OF GAUGES IN}
\centerline {\bf LINEARIZED GENERAL RELATIVITY}
\vskip 1cm
\leftline {Giampiero Esposito and Cosimo Stornaiolo}
\vskip 0.3cm
\noindent
{\it Istituto Nazionale di Fisica Nucleare, Sezione di Napoli,
Mostra d'Oltremare Padiglione 20, 80125 Napoli, Italy.}
\vskip 0.3cm
\noindent
{\it Universit\`a di Napoli Federico II, Dipartimento di Scienze
Fisiche, Complesso Universitario di Monte S. Angelo, Via Cintia,
Edificio G, 80126 Napoli, Italy.}
\vskip 1cm
\noindent
{\bf Abstract.}
For vacuum Maxwell theory in four dimensions, a supplementary
condition exists (due to Eastwood and Singer) which is invariant
under conformal rescalings of the metric, in agreement with the
conformal symmetry of the Maxwell equations. Thus, starting from
the de Donder gauge, which is not conformally invariant but is the
gravitational counterpart of the Lorenz gauge, one can consider,
led by formal analogy, a new family of gauges in general relativity,
which involve fifth-order covariant derivatives of metric 
perturbations. The admissibility of such gauges in the classical 
theory is first proven in the cases of linearized theory about flat
Euclidean space or flat Minkowski space-time. In the former, the
general solution of the equation for the fulfillment of the gauge
condition after infinitesimal diffeomorphisms involves a 3-harmonic
1-form and an inverse Fourier transform. In
the latter, one needs instead the kernel of powers of the wave 
operator, and a contour integral. The analysis is also used 
to put restrictions on the dimensionless parameter occurring in
the DeWitt supermetric, while the proof of admissibility 
is generalized to a suitable class of curved Riemannian backgrounds.
Eventually, a non-local construction is obtained of the tensor 
field which makes it possible to achieve conformal invariance
of the above gauges.
\vskip 100cm
\leftline {\bf 1. Introduction}
\vskip 0.3cm
\noindent
The transformation properties of classical and quantum field 
theories under conformal rescalings of the metric
have led, over the years, to many deep developments in 
mathematics and theoretical physics, e.g. conformal-infinity
techniques in general relativity,$^{1}$ twistor methods for gravitation
and Yang--Mills theory,$^{2,3}$ the conformal-variation method in 
heat-kernel asymptotics,$^{4}$ the discovery of conformal anomalies 
in quantum field theory.$^{5}$ All these topics are quite relevant for 
the analysis of theories which possess a gauge freedom. As a first
example, one may consider the simplest gauge theory, i.e. vacuum
Maxwell theory in four dimensions in the absence of sources.
At the classical level, the operator acting on the potential $A_{b}$
is found to be
$$
P_{a}^{\; b}=-\delta_{a}^{\; b}\cstok{\ }
+R_{a}^{\; b}+\nabla_{a}\nabla^{b},
\eqno (1.1)
$$
where $\nabla$ is the Levi--Civita connection on space-time,
$\cstok{\ } \equiv g^{ab} \nabla_{a} \nabla_{b}$, and $R^{ab}$
is the Ricci tensor. Thus, the supplementary (or gauge) condition
of the Lorenz type, i.e.
$$
\nabla^{b}A_{b}=0,
\eqno (1.2a)
$$
is of crucial importance to obtain a wave equation for $A_{b}$.
The drawback of Eq. (1.2a), however, is that it is not preserved
under conformal rescalings of the metric:
$$
{\widehat g}_{ab}=\Omega^{2} g_{ab}, \; \; \; \; 
{\widehat g}^{ab}=\Omega^{-2} g^{ab},
\eqno (1.3)
$$
whereas the Maxwell equations
$$
\nabla^{b}F_{ab}=0
\eqno (1.4)
$$
are invariant under the rescalings (1.3). This remark was the
starting point of the investigation 
by Eastwood and Singer,$^{6}$ who
found that a conformally invariant supplementary condition
may be imposed, i.e.
$$
\nabla_{b} \left[\Bigr(\nabla^{b}\nabla^{a}-2R^{ab}
+{2\over 3}R g^{ab} \Bigr)A_{a} \right]=0.
\eqno (1.5a)
$$
As is clear from Eq. (1.5a), conformal invariance is achieved at the
price of introducing third-order derivatives of the potential.
In flat backgrounds, such a condition reduces to
$$
\cstok{\ }\nabla^{b}A_{b}=0.
\eqno (1.6)
$$
Of course, all solutions of the Lorenz gauge are also solutions
of Eq. (1.6), whereas the converse does not hold.

Leaving aside the severe technical problems resulting from the attempt 
to quantize in the Eastwood--Singer gauge,$^{7}$ we are now interested
in understanding the key features of the counterpart for Einstein's
theory of general relativity. In other words, although the vacuum
Einstein equations
$$
R_{ab}-{1\over 2}g_{ab}R=0
\eqno (1.7)
$$
are not invariant under the conformal rescalings (1.3), we would like
to see whether the geometric structures leading to Eq. (1.5a) admit
a non-trivial generalization to Einstein's theory, so that a
conformally invariant supplementary condition with a higher order
operator may be found as well. For this purpose, we re-express 
Eqs. (1.2a) and (1.5a) in the form
$$
g^{ab}\nabla_{a}A_{b}=0,
\eqno (1.2b)
$$
$$ 
\eqalignno{
\; & g^{ab}\nabla_{a}\nabla_{b}\nabla^{c}A_{c}
+\left[\nabla_{b}\Bigr(-2R^{ba}+{2\over 3}R g^{ba} \Bigr)
\right]A_{a} \cr
&+ \left(-2R^{ba}+{2\over 3}R g^{ba} \right)
\nabla_{b}A_{a}=0.
&(1.5b)\cr}
$$
Eq. (1.2b) involves the space-time metric in its contravariant
form, which is also the metric on the bundle of 1-forms.
In Einstein's theory, one deals instead with the vector bundle of 
symmetric rank-two tensors on space-time
with DeWitt supermetric
$$
E^{abcd} \equiv {1\over 2} \Bigr(g^{ac}g^{bd}
+g^{ad}g^{bc}+ \alpha g^{ab}g^{cd} \Bigr),
\eqno (1.8)
$$
$\alpha$ being a real parameter different from $-{2\over m}$,
where $m$ is the dimension of space-time (this restriction on
$\alpha$ is necessary to make sure that the metric $E^{abcd}$
has an inverse). One is thus led to replace
Eq. (1.2b) with the de Donder gauge
$$
W^{a} \equiv E^{abcd}\nabla_{b}h_{cd}=0.
\eqno (1.9)
$$
Hereafter, $h_{ab}$ denotes metric perturbations, since we are
interested in linearized general relativity.
The supplementary condition (1.9) is not invariant under conformal
rescalings, but the expression of the Eastwood--Singer gauge in
the form (1.5b) suggests considering as a ``candidate'' for a
conformally invariant gauge involving a higher-order operator
the equation
$$
E^{abcd}\nabla_{a}\nabla_{b}\nabla_{c}\nabla_{d}W^{e}
+\left[\Bigr(\nabla_{p}T^{pebc}\Bigr)+T^{pebc}\nabla_{p}
\right]h_{bc}=0 .
\eqno (1.10)
$$
More precisely, Eq. (1.10) is obtained from Eq. (1.5b) by applying
the replacement prescriptions 
$$
g^{ab} \rightarrow E^{abcd},
\; A_{b} \rightarrow h_{ab}, 
\; \nabla^{b}A_{b} \rightarrow W^{e},
$$ 
with $T^{pebc}$ a rank-four tensor field obtained from
the Riemann tensor, the Ricci tensor, the trace of Ricci and
the metric. In other words, $T^{pebc}$ is expected to include
all possible contributions of the kind $R^{pebc}, R^{pe} g^{bc},
R g^{pe} g^{bc}$. We will however see in Sec. 5 that 
$T^{pebc}$ is even more involved.

When a supplementary (or gauge) condition is imposed in a theory
with gauge freedom, one of the first problems is to make sure that
such a condition is preserved under the action of the gauge 
symmetry. More precisely, either the gauge is originally satisfied,
and hence also the gauge-equivalent field configuration should fulfill
the condition, or the gauge is not originally satisfied, but one wants 
to prove that, after performing a gauge transformation, it is always
possible to fulfill the supplementary condition, eventually. The
latter problem is the most general, and has a well known counterpart
already for Maxwell theory (see Sec. 6.5 
of Ref. 8). For linearized classical
general relativity in the family of gauges described by Eq. (1.10),
the gauge symmetry remains the request of invariance under infinitesimal
diffeomorphisms. Their effect on metric perturbations is given by
$$
{ }^{\varphi}h_{ab} \equiv h_{ab}+(L_{\varphi}h)_{ab}
=h_{ab}+\nabla_{(a} \; \varphi_{b)}.
\eqno (1.11)
$$
For some smooth metric perturbation one might indeed have
(cf. Eq. (1.10))
$$
E^{abcd}\nabla_{a}\nabla_{b}\nabla_{c}\nabla_{d}W^{e}(h)
+\Bigr[(\nabla_{p}T^{pebc})+T^{pebc}\nabla_{p}\Bigr]h_{bc}
\not = 0.
\eqno (1.12)
$$
We would like to prove that one can, nevertheless, 
achieve the condition
$$
E^{abcd}\nabla_{a}\nabla_{b}\nabla_{c}\nabla_{d}
W^{e}({ }^{\varphi}h)+\Bigr[(\nabla_{p}T^{pebc})
+T^{pebc}\nabla_{p}\Bigr]{ }^{\varphi}h_{bc}=0.
\eqno (1.13)
$$
Equation (1.13) is conveniently re-expressed in a form where the
left-hand side involves a differential operator acting on the
1-form $\varphi_{q}$, and the right-hand side depends only on
metric perturbations, their covariant derivatives and the Riemann
curvature. Explicitly, one finds
$$
P_{e}^{\; q} \; \varphi_{q}=-F_{e},
\eqno (1.14)
$$
where (hereafter $h$ is the trace $g^{ab}h_{ab}$)
$$ \eqalignno{
\; & P_{e}^{\; q} \equiv \left(\nabla^{(c} \; \nabla^{d)}
\nabla_{c}\nabla_{d}+{\alpha \over 2} \cstok{\ }^{2}\right)
\left(\delta_{e}^{\; q} \cstok{\ }+\nabla^{q}\nabla_{e}
+\alpha \nabla_{e}\nabla^{q}\right) \cr
&+2 T_{\; \; e \; \; \; \; \; ;p}^{p \; (bq)} \;
\nabla_{b}+2T_{\; \; e}^{p \; (bq)} \nabla_{p}\nabla_{b},
&(1.15)\cr}
$$
$$ \eqalignno{
\; & F_{e} \equiv 2 \left(\nabla^{(c} \; \nabla^{d)}
\nabla_{c}\nabla_{d}+{\alpha \over 2}\cstok{\ }^{2}\right)
\left(\nabla^{q}h_{qe}+{\alpha \over 2}\nabla_{e}h \right) \cr
&+2 T_{\; \; e \; \; \; \; ;p}^{p \; \; bc} \; h_{bc}
+2 T_{\; \; e}^{p \; \; bc} \; \nabla_{p}h_{bc}.
&(1.16)\cr}
$$

Section 2 solves Eq. (1.14) via Fourier transform when the
Riemann curvature of the background vanishes and the metric $g$
is positive-definite. Linearized theory about $m$-dimensional
Minkowski space-time is studied in Sec. 3. Section 4 shows how
to solve Eq. (1.14) in curved Riemannian backgrounds without
boundary, and the construction of conformally invariant gauges
is obtained in Sec. 5 in non-local form. 
Concluding remarks are presented in Sec. 6.
\vskip 0.3cm
\leftline {\bf 2. Linearized Theory about Flat Euclidean Space}
\vskip 0.3cm
\noindent
It may be helpful to begin the analysis of Eq. (1.14) in the
limiting case when the Riemann curvature of the background geometry
$(M,g)$ vanishes. This means that one is considering linearized
theory about flat space-time, or flat space if $g$ is taken to be
positive-definite. It remains useful, however, to use a notation
in terms of covariant derivatives in $P_{e}^{\; q}$ and $F_{e}$, not
only to achieve covariance, but also because the flat background
might have a curved boundary $\partial M$, and hence $\nabla_{c}$
might be re-expressed in terms of covariant derivatives with respect
to the induced connection on $\partial M$, after taking into account
the extrinsic curvature of $\partial M$.

Under the above assumptions, Eq. (1.14) becomes a partial differential
equation involving a sixth-order differential operator with constant
coefficients. It is therefore convenient to take the Fourier transform
(denoted by a tilde) of both sides of Eq. (1.14), because it is well
known that the Fourier transform turns a constant coefficient 
differential operator into a multiplication operator. Bearing in
mind the definitions (hereafter ${\bf E}^{m}$ is flat
$m$-dimensional Euclidean space)
$$
{\widetilde H}_{q}(\xi) \equiv (2\pi)^{-{m\over 2}} 
\int_{{\bf E}^{m}} H_{q}(x)e^{-i \xi_{a}x^{a}} dx^{1}...dx^{m},
\eqno (2.1)
$$
$$
H_{q}(x) \equiv (2\pi)^{-{m\over 2}}
\int_{{\bf E}^{m}}{\widetilde H}_{q}(\xi)e^{i \xi_{a}x^{a}}
d\xi_{1}...d\xi_{m},
\eqno (2.2)
$$
where $H_{q}$ may be $\varphi_{q}$ or $F_{q}$, and 
$\xi \in T^{*}(M)$, one then finds the equation
$$
\sigma[P_{e}^{\; q}](\xi){\widetilde \varphi}_{q}(\xi)
=-{\widetilde F}_{e}(\xi),
\eqno (2.3)
$$
where $\sigma[P_{e}^{\; q}](\xi)$ is the symbol$^{4}$ of the operator
$P_{e}^{\; q}$, obtained by replacing $\nabla_{a}$ with $i\xi_{a}$. 
In other words, we insert Eq. (2.2) and
the first lines of Eqs. (1.15) and (1.16) into Eq. (1.14). This leads
to Eq. (2.3), where
$$
\sigma[P_{e}^{\; q}](\xi)=-\left(1+{\alpha \over 2}\right)
(\xi_{a}\xi^{a})^{3}\left[\delta_{e}^{\; q}
+(1+\alpha){\xi_{e}\xi^{q}\over \xi_{a}\xi^{a}}\right].
\eqno (2.4)
$$
Our aim is now to solve Eq. (2.3) for 
${\widetilde \varphi}_{q}(\xi)$, and eventually anti-transform to
get $\varphi_{q}(x)$. For this purpose, we first study a
background metric $g$ which is Riemannian. This implies that
$\xi_{a}\xi^{a}=g(\xi,\xi) \not = 0$ for all $\xi \not =0$,
and hence one has to rule out the value $\alpha=-2$ of the parameter
$\alpha$ in the supermetric (1.8) to obtain a well defined inverse
of $\sigma[P_{e}^{\; q}]$. Under these assumptions one finds, after
setting
$$
\rho_{e}^{\; q}(\xi) \equiv \sigma[P_{e}^{\; q}](\xi),
\eqno (2.5)
$$
that the inverse of the symbol is
$$
{\rho^{-1}}_{e}^{\; q}=-{\left(1+{\alpha \over 2}\right)}^{-1}
(\xi_{a}\xi^{a})^{-3}\left[\delta_{e}^{\; q}
-{(1+\alpha)\over (2+\alpha)}
{\xi_{e}\xi^{q}\over \xi_{a}\xi^{a}}\right].
\eqno (2.6)
$$
Thus, equation (2.3) can be solved for the Fourier transform of
$\varphi_{q}$ as
$$
{\widetilde \varphi}_{q}(\xi)=
{\left(1+{\alpha \over 2}\right)}^{-1}
(\xi_{a}\xi^{a})^{-3}\left[\delta_{q}^{\; s}
-{(1+\alpha)\over (2+\alpha)}
{\xi_{q}\xi^{s}\over \xi_{a}\xi^{a}}\right]
{\widetilde F}_{s}(\xi),
\eqno (2.7)
$$
for all $\xi \not = 0$.
This leads, by inverse Fourier transform, to the desired formula 
for the 1-form $\varphi_{q}(x)$, i.e. (see (1.16) and (2.2))
$$
\varphi_{q}(x)={(2\pi)^{-{m\over 2}}\over 
\left(1+{\alpha \over 2}\right)} 
\int_{{\bf E}^{m}}
|\xi|^{-6}
\left[\delta_{q}^{\; s}
-{(1+\alpha)\over (2+\alpha)}
{\xi_{q}\xi^{s}\over |\xi|^{2}}\right]
{\widetilde F}_{s}(\xi) e^{i\xi \cdot x } d\xi,
\eqno (2.8)
$$
where we have defined $|\xi|^{2} \equiv \xi_{a}\xi^{a}$,
$\xi \cdot x \equiv \xi_{a}x^{a}$
and $d\xi \equiv d\xi_{1}...d\xi_{m}$. The only poles of the
integrand occur when 
$$
\xi_{0}=\pm i \sqrt{\sum_{k=1}^{m-1}\xi_{k}\xi^{k}},
$$
i.e. on the imaginary $\xi_{0}$ axis. Thus, integration on the
real line for $\xi_{0}$, and subsequent integration with respect
to $\xi_{1},...,\xi_{m-1}$, yields a well defined integral
representation of $\varphi_{q}$. This is entirely analogous to
the integration performed in quantum field theory to define the
Euclidean form of the Feynman Green function (see section 2.7
of Ref. 9, first item).

In particular, when $\alpha=-1$, the operator on $\varphi_{q}$ 
reduces to the cubic Laplacian, i.e. the Laplacian composed twice 
with itself:
$$
\cstok{\ } \; \cstok{\ } \; \cstok{\ }
=g^{ab}g^{cd}g^{ef}\nabla_{a}\nabla_{b}\nabla_{c}\nabla_{d}
\nabla_{e}\nabla_{f}.
$$
Note, however, that Eq. (2.8) does not yield the general solution
of Eq. (1.14). For this purpose, one has to add to (2.8) the general
solution of the homogeneous equation
$$
P_{a}^{\; b}\varphi_{b}(x)=0.
\eqno (2.9)
$$
In particular, if $\alpha$ is set equal to $-1$, Eqs. (1.15) and
(2.9) lead to
$$
\cstok{\ }^{3}\varphi_{a}(x)=0 \; \; \; \forall \; a=1,...,m.
\eqno (2.10)
$$
Since we are considering flat space in Cartesian coordinates, as
is clear already from the definitions (2.1) and (2.2), Eq. (2.10)
reads, explicitly,
$$
g^{bc}g^{de}g^{fh}{\partial^{6}\varphi_{a}(x) \over
\partial x^{b} \partial x^{c} \partial x^{d} \partial x^{e}
\partial x^{f} \partial x^{h}}=0
\; \; \; \forall \; a=1,...,m.
\eqno (2.11)
$$
Thus, all components of $\varphi_{a}$ should be represented by
3-harmonic functions, which, by definition, satisfy the equation
$$
\cstok{\ }^{3}f(x)=0.
\eqno (2.12)
$$
What we need is the following structural property:
\vskip 0.3cm
\noindent
{\bf Theorem 2.1.} Every 3-harmonic function $f$ in ${\bf E}^{m}$ is
completely determined by three harmonic functions and by the Green
kernel of the Laplacian.
\vskip 0.3cm
\noindent
{\bf Proof.} Let us define the functions $u$ and $v$ by the equations
$$
u(x) \equiv \cstok{\ }f(x),
\eqno (2.13)
$$
$$
v(x) \equiv \cstok{\ }u(x).
\eqno (2.14)
$$
We then find, by virtue of Eq. (2.12), that $v$ is harmonic:
$$
0=\cstok{\ }^{3}f(x)=\cstok{\ }^{2}u(x)=\cstok{\ }v(x).
\eqno (2.15)
$$
We now use the Green kernel $G(x,y)$ of the Laplacian, for 
which ($\delta(x,y)$ being the Dirac distribution)
$$
\cstok{\ }_{x}G(x,y)=\delta(x,y),
\eqno (2.16)
$$
where the subscript for the Laplacian is used to denote its
action as a differential operator on the coordinates of the
first argument of the kernel. The function $u$ can be then
expressed as ($dy=dy_{1}...dy_{m}$ being the integration
measure on ${\bf E}^{m}$)
$$
u(x)=\cstok{\ }^{-1}v(x)=\int_{{\bf E}^{m}}G(x,y)v(y)dy,
\eqno (2.17)
$$
and hence, from (2.13),
$$ \eqalignno{
\; & f(x)=\cstok{\ }^{-1}u(x)=\int_{{\bf E}^{m}}G(x,y)u(y)dy \cr
&=\int_{{\bf E}^{m}} \int_{{\bf E}^{m}}
G(x,y)G(y,z)v(z)dy \; dz.
&(2.18)\cr}
$$
This is not, however, the most general solution of Eq. (2.12). One
can in fact add to the integral (2.18) a harmonic function $f_{1}$
and a bi-harmonic function $f_{2}$, because, if $h$ is a
$n$-harmonic function, for which ($n$ being $\geq 1$)
$$
\cstok{\ }^{n}h(x)=0 ,
\eqno (2.19)
$$
then $h$ is also $(n+m)$-harmonic, with $m \geq 1$, whereas the
converse does not necessarily hold. 
In our case, since $n=3$, this implies that
the general solution of Eq. (2.12) can be written as
$$
f(x)=f_{1}(x)+f_{2}(x)+\int_{{\bf E}^{m}} \int_{{\bf E}^{m}}
G(x,y) G(y,z) v(z) dy \; dz ,
\eqno (2.20)
$$
where $f_{1}$ is harmonic and $f_{2}$ is bi-harmonic:
$$
\cstok{\ }f_{1}(x)=0,
\eqno (2.21)
$$
$$
\cstok{\ }^{2}f_{2}(x)=0.
\eqno (2.22)
$$
We now apply the same procedure to $f_{2}$, to write it as
$$
f_{2}(x)=g_{1}(x)+\int_{{\bf E}^{m}}G(x,y)w(y)dy,
\eqno (2.23)
$$
where $g_{1}$ and $w$ are harmonic. By virtue of (2.20) and (2.23)
we can write
$$ \eqalignno{
\; & f(x)=\Omega(x)+\int_{{\bf E}^{m}}G(x,y)w(y)dy \cr
&+\int_{{\bf E}^{m}} \int_{{\bf E}^{m}}G(x,y)G(y,z)v(z)
dy \; dz ,
&(2.24)\cr}
$$
where $\Omega$ is the harmonic function equal to $f_{1}+g_{1}$. This
completes the proof of the result we needed (cf. Ref. 10).

The general solution of Eq. (1.14) reads therefore,
when $\alpha=-1$ (see (2.8)),
$$ \eqalignno{
\; & \varphi_{a}(x)=\Omega_{a}(x)+\int_{{\bf E}^{m}}
G_{a}^{\; b}(x,y)w_{b}(y)dy \cr
&+\int_{{\bf E}^{m}} \int_{{\bf E}^{m}}
G_{a}^{\; b}(x,y) G_{b}^{\; c}(y,z)v_{c}(z)dy \; dz \cr
&+2(2\pi)^{-{m\over 2}}\int_{{\bf E}^{m}}
|\xi|^{-6} {\widetilde F}_{a}(\xi)
e^{i \xi \cdot x} d\xi,
&(2.25)\cr}
$$
where $\Omega_{a},w_{a}$ and $v_{a}$ are harmonic 1-forms 
in ${\bf E}^{m}$.

In the applications, it may be useful to consider the Euclidean
4-ball,$^{11}$ i.e. a portion of flat Euclidean 4-space bounded by a
3-sphere of radius $a$. The cubic Laplacian on normal and tangential
components of $\varphi_{a}$ leads to sixth-order operators with a
regular singular point at $\tau=0$, where $\tau$ is the radial
coordinate lying in the closed interval $[0,a]$. After defining
the new independent variable $T \equiv \log(\tau)$, the differential
equations involving such operators are mapped into equations
with constant coefficient operators (see Ref. 11 for the
$\cstok{\ }$ and $\cstok{\ }^{2}$ operators), and hence one can use
again Fourier transform techniques to find a particular solution of
Eq. (1.14). The general structure of $\varphi_{a}(x)$ is therefore
elucidated quite well by the result (2.25) in Cartesian coordinates.
Strictly speaking, we should of course replace $\varphi_{a}$ by
$\varphi_{a}dx^{a}$, since only the latter represents a 1-form. 
The same holds for all the other 1-forms considered hereafter.
\vskip 10cm
\leftline {\bf 3. Linearized Theory About Flat Space-Time}
\vskip 0.3cm
\noindent
If the background metric is Minkowskian, one can first define
the partial Fourier transform with
respect to the time variable, according to the rule
$$
{\widetilde \varphi}_{a}({\vec x},\omega) \equiv
{1\over \sqrt{2\pi}} \int_{-\infty}^{\infty}
\varphi_{a}({\vec x},t)e^{i \omega t}dt ,
\eqno (3.1)
$$
where ${\vec x} \equiv (x^{1},...,x^{m-1})$, with the 
corresponding anti-transform 
$$
\varphi_{a}({\vec x},t) \equiv {1\over \sqrt{2\pi}}
\int_{-\infty}^{\infty} {\widetilde \varphi}_{a}({\vec x},\omega)
e^{-i \omega t} d\omega.
\eqno (3.2)
$$
The operator $P_{a}^{\; b}$ is studied in $m$-dimensional Minkowski
space-time and, for simplicity, we set $\alpha=-1$
(it remains necessary to rule out $\alpha=-2$ to obtain a
meaningful solution). Equation (1.14) becomes, therefore,
$$
\cstok{\ }^{3}\varphi_{a}({\vec x},t)=-2F_{a}({\vec x},t),
\eqno (3.3)
$$
where $\cstok{\ }$ is now the wave operator, i.e.
(here $x^{0}=ct$)
$$
\cstok{\ } \equiv -{1\over c^{2}}{\partial^{2}\over \partial t^{2}}
+\sum_{i,j=1}^{m-1}g^{ij} {\partial\over \partial x^{i}}
{\partial \over \partial x^{j}}
\equiv -{1\over c^{2}}{\partial^{2}\over \partial t^{2}}
+\bigtriangleup .
\eqno (3.4)
$$
By virtue of (3.2) and (3.4), Eq. (3.3) leads to the following
equation for the partial Fourier transform 
${\widetilde \varphi}_{a}({\vec x},\omega)$: 
$$
\left[\bigtriangleup^{3}+3{\omega^{2}\over c^{2}}
\bigtriangleup^{2}+3{\omega^{4}\over c^{4}}\bigtriangleup
+{\omega^{6}\over c^{6}}\right]
{\widetilde \varphi}_{a}({\vec x},\omega)
=-2 {\widetilde F}_{a}({\vec x},\omega).
\eqno (3.5)
$$
One can now use a Green-function approach, and look for the Green
function of the operator on the left-hand side of (3.5). On setting
$R \equiv |{\vec x}-{\vec x}'|$, $k \equiv {\omega \over c}$, this
is a solution, for all $R \not = 0$, of the equation
$$
\left[\bigtriangleup^{3}+3k^{2}\bigtriangleup^{2}
+3k^{4}\bigtriangleup+k^{6} \right]G_{k}(R)=0,
\eqno (3.6)
$$
where
$$
\bigtriangleup={d^{2}\over dR^{2}}+{(m-1)\over R}{d\over dR}.
\eqno (3.7)
$$
The desired Green function can be factorized in the form 
$$
G_{k}(R)=R^{\delta}F(R),
\eqno (3.8)
$$
and the parameter $\delta$ is eventually found to depend on $m$, 
to obtain a simplified form of the sixth-order differential
operator under investigation. For example, for the Helmholtz
equation
$$
(\bigtriangleup +k^{2})G_{k}(R)=0 \; \; \; \forall R \not = 0,
$$
associated to the wave equation, this method yields
$$
G_{k}(R)=r^{-{(m-1)\over 2}}e^{\pm ikR}.
$$
In our case, however, Eqs. (3.6) and (3.7) lead to a lengthy
calculation, whereas the partial Fourier transform of Eq. (3.5)
with respect to the spatial variables leads more quickly to the
Lorentzian counterpart of Eq. (2.8). In other words, it is
convenient to define (here ${\vec \xi} \equiv (\xi_{1},...,\xi_{m-1})$)
$$
{\widehat \varphi}_{a}({\vec \xi},\omega) \equiv
(2\pi)^{-{(m-1)\over 2}} \int_{{\bf R}^{m-1}}
{\widetilde \varphi}_{a}({\vec x},\omega)e^{-i \xi_{k}x^{k}}
dx^{1}...dx^{m-1},
\eqno (3.9)
$$
and the associated Fourier anti-transform
$$
{\widetilde \varphi}_{a}({\vec x},\omega) \equiv
(2\pi)^{-{(m-1)\over 2}} \int_{{\bf R}^{m-1}}
{\widehat \varphi}_{a}({\vec \xi},\omega)
e^{i \xi_{k} x^{k}} d\xi_{1} ... d\xi_{m-1}.
\eqno (3.10)
$$
Equations (3.5) and (3.10) lead to 
$$
\left[-|{\vec \xi}|^{6}+3\xi_{0}^{2}|{\vec \xi}|^{4}
-3\xi_{0}^{4}|{\vec \xi}|^{2}+\xi_{0}^{6}\right]
{\widehat \varphi}_{a}({\vec \xi},\omega)
=-2{\widehat F}_{a}({\vec \xi},\omega),
\eqno (3.11)
$$
where $|{\vec \xi}|^{2} \equiv \sum_{i,j=1}^{m-1}g^{ij}
\xi_{i} \xi_{j}$, $\xi_{0} ={\omega \over c}$. Some care
is necessary to invert this equation, because the polynomial on
the left-hand side is equal to $-\left[ |{\vec \xi}|^{2}-\xi_{0}^{2}
\right]^{3}$, which vanishes if
$$
\xi_{0}=-\sqrt{|{\vec \xi}|^{2}}=\xi_{0,1},
\eqno (3.12)
$$
or 
$$
\xi_{0}=+\sqrt{|{\vec \xi}|^{2}}=\xi_{0,2}.
\eqno (3.13)
$$
This means that the $\xi_{0}$ integration is singular and there are
two poles on the real $\xi_{0}$ axis at the points given by (3.12)
and (3.13). In other words, the integration with respect to $\xi_{0}$
is a contour integral which may be performed by deforming the 
contour around the poles. The way in which this deformation is
performed determines the particular solution of Eq. (3.3).
To deal with the poles, we choose a Feynman-type contour
that comes from $-\infty-i \varepsilon$, encircles the points
$\xi_{0,1}$ and $\xi_{0,2}$ from below and from the above, 
respectively, and goes off to $+\infty+i \varepsilon$. This operation
is denoted with the standard $+i \varepsilon$ symbol, with a contour
that we call $\gamma_{m}$. Thus,
by virtue of (3.2), (3.10) and (3.11) we find, if
$\alpha=-1$, a particular solution of Eq. (1.14) in the form (cf.
the third line of (2.25))
$$
\varphi_{a}({\vec x},t)=2(2\pi)^{-{m\over 2}}
\int_{\gamma_{m}}[(|{\vec \xi}|^{2}-\xi_{0}^{2})^{3}
+i \varepsilon]^{-1}{\widehat F}_{a}({\vec \xi},\omega)
e^{ig^{cd}\xi_{c}x_{d}}d\xi.
\eqno (3.14)
$$
With this understanding, the complete solution of
Eq. (1.14) when $\alpha=-1$ is formally similar to Eq. (2.25), but
bearing in mind that $G_{a}^{\; b}({\vec x},t;{\vec y},t')$ is now
the Green kernel of the wave operator, and that $\Omega_{a}, w_{a}$
and $v_{a}$ are three solutions of the homogeneous wave 
equation $\cstok{\ }A_{a}=0$. Unlike the case of Euclidean background,
various boundary conditions, and hence various choices of contour,
lead to different solutions. 
\vskip 0.3cm
\leftline {\bf 4. Generalization to Curved Backgrounds}
\vskip 0.3cm
\noindent
If one studies linearized gravity about curved backgrounds $M$, it is
no longer possible to express the Fourier transform of Eq. (1.14) 
in the form (2.3), because the second line of Eq. (1.15) contributes
terms which cannot be factorized under Fourier transform. 
However, it remains true that the operator $P_{e}^{\; q}$ is the
sum of a sixth-order differential operator ${\cal A}_{e}^{\; q}$ with
constant coefficients and other terms 
$$
{\cal B}_{e}^{\; q} \equiv
2 T_{\; \; e \; \; \; \; \; ;p}^{p \; (bq)} \; \nabla_{b}
+2T_{\; \; e}^{p \; (bq)} \nabla_{p}\nabla_{b},
\eqno (4.1)
$$
involving curvature and lower-order covariant derivatives. We are
therefore studying the partial differential equation
$$
\Bigr({\cal A}_{e}^{\; q}+{\cal B}_{e}^{\; q}\Bigr)
\varphi_{q}(x)=-F_{e}(x).
\eqno (4.2)
$$
Hereafter we focus on Riemannian backgrounds (i.e. with
positive-definite metric) so as to be able to use some well
established results which rely on the theory of elliptic
operators on compact Riemannian manifolds without boundary.$^{4}$
By construction, the symbol of ${\cal A}_{e}^{\; q}$ coincides
with the expression (2.4), and hence the operator 
${\cal A}_{e}^{\; q}$ is invertible if and only if $\alpha \not 
= -2$. Let ${\cal G}_{e}^{\; q}(x,y)$ denote the corresponding Green
kernel. The application of the inverse operator ${\cal A}^{-1}$
to both sides of Eq. (4.2) yields therefore the integral equation
$$
\varphi_{e}(x)+\int_{M} {\cal G}_{e}^{\; p}(x,y)
\Bigr({\cal B}_{p}^{\; r}
\varphi_{r}\Bigr)(y)\sqrt{g(y)}dy
+\int_{M} {\cal G}_{e}^{\; p}(x,y)F_{p}(y)\sqrt{g(y)}dy=0,
\eqno (4.3)
$$
where $g(y)$ denotes the determinant of the background metric.
To solve for $\varphi_{e}(x)$ we have to study the action of
${\cal B}_{p}^{\; r}$ on
$\varphi_{r}$. For this purpose, we assume that the curvature of
the background is such that ${\cal B}_{p}^{\; r}$ is a symmetric
elliptic operator, so that a discrete spectral resolution with
eigenvectors of class $C^{\infty}$ exists (see Sec. 1.6 of Ref. 4).
The eigenvectors $\varphi_{r}^{(n)}(x)$, for which
($\lambda_{(n)}$ denoting the eigenvalues)
$$
{\cal B}_{p}^{\; r}\varphi_{r}^{(n)}(x)=\lambda_{(n)}
\varphi_{p}^{(n)}(x),
\eqno (4.4)
$$
make it then possible to expand $\varphi_{r}(x)$ in the form
$$
\varphi_{r}(x)=\sum_{n=1}^{\infty}a_{(n)}\varphi_{r}^{(n)}(x)
+{\overline \varphi}_{r}(x),
\eqno (4.5)
$$
where ${\overline \varphi}_{r}(x) \in {\rm Ker} {\cal B}_{p}^{\; r}$,
i.e. ${\cal B}_{p}^{\; r}{\overline \varphi}_{r}(x)=0$.
Note that the ellipticity condition means that the leading symbol
of ${\cal B}_{p}^{\; r}$ is non-vanishing for all
$\xi_{a} \not =0$, i.e.
$$
-2T_{\; \; e}^{p \; (bq)} \xi_{p}\xi_{b} \not = 0
\; {\rm for} \; \xi \not = 0.
\eqno (4.6)
$$
This condition receives contributions from the parts of $T$ involving
the Ricci tensor and the scalar curvature, but not from the Riemann
tensor, which is antisymmetric in $b$ and $q$. For a given choice
of background with associated curvature and tensor $T$, the above
condition provides a useful operational criterion to check the
admissibility of our supplementary condition (1.10). If the leading
symbol of ${\cal B}_{p}^{\; r}$ 
is further taken to be positive-definite for
$\xi \not =0$, so as to ensure that the spectrum 
of ${\cal B}_{p}^{\; r}$ is
bounded from below,$^{4}$ one gets another useful operational
criterion expressed by a majorization, i.e (cf. (4.6))
$$
-T_{\; \; e}^{p \; (bq)} \xi_{p}\xi_{b} > 0
\; {\rm for} \; \xi \not = 0.
\eqno (4.7)
$$

By virtue of the above assumptions, on defining
$$
f_{r}(x)\equiv \int_{M} {\cal G}_{r}^{\; q}(x,y)F_{q}(y)\sqrt{g(y)}dy,
\eqno (4.8)
$$
its expansion involves a set of coefficients $f_{(n)}$ (replacing
$a_{(n)}$), and possibly a remainder term 
${\overline f}_{r}(x)$ (replacing
${\overline \varphi}_{r}(x)$) in the kernel of ${\cal B}_{p}^{\; r}$:
$$
f_{r}(x)=\sum_{n=1}^{\infty}f_{(n)}\varphi_{r}^{(n)}(x)
+{\overline f}_{r}(x).
\eqno (4.9)
$$
Moreover, the insertion of (4.5) into Eq. (4.3) leads to the
analysis of $1$-form-valued ``coefficients'' $\gamma_{r}^{(m)}(x)$
defined by
$$
\gamma_{r}^{(m)}(x) \equiv \int_{M} 
{\cal G}_{r}^{\; p}(x,y)\varphi_{p}^{(m)}(y)\sqrt{g(y)}dy,
\eqno (4.10)
$$
which can be expanded in the form (cf. (4.5))
$$
\gamma_{r}^{(m)}(x)=\sum_{n=1}^{\infty}
\gamma^{(m)(n)}\varphi_{r}^{(n)}(x)
+{\overline \gamma}_{r}^{(m)}(x),
\eqno (4.11)
$$
where ${\overline \gamma}_{r}^{(m)}(x)$ belongs to the kernel of
${\cal B}_{p}^{\; r}$. By virtue of 
(4.4)--(4.11), Eq. (4.3) may be re-expressed in the form
$$ \eqalignno{
\; & \sum_{n=1}^{\infty}\Bigr(a_{(n)}+\sum_{m=1}^{\infty}
\lambda_{(m)}a_{(m)}\gamma^{(m)(n)}+f_{(n)}\Bigr)\varphi_{r}^{(n)}(x) \cr
&+{\overline \varphi}_{r}(x)+{\overline f}_{r}(x)
+\sum_{m=1}^{\infty}\lambda_{(m)}a_{(m)}
{\overline \gamma}_{r}^{(m)}(x)=0.
&(4.12)\cr}
$$
The first and second line of Eq. (4.12) should vanish separately,
and hence one finds an infinite set of equations for the 
coefficients $a_{(n)}$:
$$
\sum_{m=1}^{\infty}\Bigr(\delta_{mn}
+\lambda_{(m)}\gamma^{(m)(n)}\Bigr)
a_{(m)}=-f_{(n)},
\eqno (4.13)
$$
jointly with the equation
$$
{\overline \varphi}_{r}(x)=-{\overline f}_{r}(x)
-\sum_{m=1}^{\infty}\lambda_{(m)}a_{(m)}
{\overline \gamma}_{r}^{(m)}(x).
\eqno (4.14)
$$
Although the technical details remain elaborated, we have obtained
a complete prescription for finding $\varphi_{e}(x)$ and hence
proving the admissibility of our gauges in curved backgrounds. 
First, we solve the inhomogeneous system (4.13) once the eigenvalues
$\lambda_{(m)}$ of the operator ${\cal B}_{p}^{\; r}$ are determined 
in the given background (this is already a hard task). 
The remaining part (if any) of $\varphi_{e}(x)$, i.e. the one in the
kernel of ${\cal B}_{p}^{\; r}$, 
is then evaluated from Eq. (4.14), and eventually the full 
$\varphi_{e}(x)$ is obtained from the expansion (4.5).
\vskip 0.3cm
\leftline {\bf 5. Construction of a Conformally Invariant Gauge}
\vskip 0.3cm
\noindent
We now consider a metric 
$\gamma=\gamma_{ab}dx^{a}\otimes dx^{b}$ solving the full Einstein
equations in vacuum ($\gamma$ is therefore the ``physical
metric''):
$$
R_{ab}(\gamma)-{1\over 2}\gamma_{ab}R(\gamma)=0.
\eqno (5.1)
$$
On denoting again by $g_{ab}$ a background 
metric,$^{12}$ we study a
gauge condition linear in $\gamma$ having the form (cf. (1.10))
$$
S^{e}(\gamma) \equiv E^{abcd}(g)\nabla_{a}\nabla_{b}\nabla_{c}
\nabla_{d} W^{e}(\gamma)+\nabla_{p}{\widetilde T}^{pe}(\gamma)=0,
\eqno (5.2)
$$
where
$$
W^{e}(\gamma) \equiv E^{epqr}(g)\nabla_{p}\gamma_{qr}.
\eqno (5.3)
$$
The tensor ${\widetilde T}^{pe}(\gamma)$ 
is linear in $\gamma$ but can be more
general than the combination $T^{pebc}\gamma_{bc}$, where 
$T^{pebc}$ is obtained by all possible permutations with different
coefficients of 
$$
R^{pebc}(g), \; R^{pe}(g) g^{bc}, \;
R(g) g^{pe} g^{bc}
$$
(cf. comments following Eq. (1.10)). In other words, 
${\widetilde T}^{pe}(\gamma)$ may contain terms like
$$
R(g)\gamma^{pe}, \; 
R^{pbec}(g)\gamma_{bc}, \; R^{pe}(g)g^{bc}\gamma_{bc}, \;
R(g)g^{pe}g^{bc}\gamma_{bc},
$$
plus many other contributions where the Riemann
tensor $R_{\; bcd}^{a}$ built from the background metric is
replaced by any tensor field $F_{\; bcd}^{a}$ of type $(1,3)$
and independent of the physical metric. Upon considering the
conformal rescalings 
$$
\gamma_{ab} \rightarrow \Omega^{2}\gamma_{ab}, \;
\gamma^{ab} \rightarrow \Omega^{-2}\gamma^{ab},
$$
the terms we have written down explicitly have conformal
weights $-2,2,2,2$, respectively.
Note that the connection
$\nabla$ leads to covariant derivatives $\nabla_{a}$ with
respect to the background metric $g$, subject to the condition
$$
\nabla g=0 \Longrightarrow g_{bc;a}=0,
\eqno (5.4)
$$
whereas the covariant derivatives of the physical metric $\gamma$
with respect to the background metric $g$ do not vanish:
$\nabla \gamma \not = 0$.$^{12}$ It will be shown that
${\widetilde T}^{pe}(\gamma)$ depends also on $\nabla \gamma$
and is obtained from a non-local construction (see below). 

We now study the behaviour of $S^{e}(\gamma)$ under conformal
rescalings of the physical metric which solves Eq. (5.1).
For this purpose, it is convenient to define
$$
Q^{e}(\gamma) \equiv E^{abcd}(g)
\nabla_{a}\nabla_{b}\nabla_{c}\nabla_{d}W^{e}(\gamma).
\eqno (5.5)
$$
We therefore find that
$$
Q^{e}(\Omega^{2}\gamma)=\Omega^{2}Q^{e}(\gamma)
+U^{e}(\gamma,\nabla^{(k)} \gamma, \Omega, 
\nabla^{(k)} \Omega),
\eqno (5.6)
$$
where $\nabla^{(k)}$ denotes covariant derivative of $k$-th
order, with $k=1,2,3,4$, and $U^{e}$ is a vector given by
$$
U^{e} \equiv E^{abcd}(g)E^{epqr}(g)Z_{abcdpqr},
\eqno (5.7)
$$
having defined
$$
Z_{abcdpqr} \equiv \sum_{k=1}^{5}Z_{abcdpqr}^{(k)},
\eqno (5.8)
$$
with
$$ \eqalignno{
\; & Z_{abcdpqr}^{(1)} \equiv 2\Omega \Bigr(\Omega_{;a}
\gamma_{qr;pdcb}+\Omega_{;b}\gamma_{qr;pdca} \cr
&+\Omega_{;c}\gamma_{qr;pdba}+\Omega_{;d}
\gamma_{qr;pcba}+\Omega_{;p}\gamma_{qr;dcba}\Bigr),
&(5.9)\cr}
$$
$$ \eqalignno{
\; & Z_{abcdpqr}^{(2)} \equiv 2\biggr[\Bigr(\Omega_{;d}\Omega_{;p}
+\Omega \Omega_{;pd}\Bigr)\gamma_{qr;cba} 
+\Bigr(\Omega_{;c}\Omega_{;p}+\Omega \Omega_{;pc}\Bigr)
\gamma_{qr;dba} \cr
&+\Bigr(\Omega_{;b}\Omega_{;p}
+\Omega \Omega_{;pb}\Bigr)\gamma_{qr;dca}
+\Bigr(\Omega_{;a}\Omega_{;p}+\Omega \Omega_{;pa}\Bigr)
\gamma_{qr;dcb} \cr
&+ \Bigr(\Omega_{;c}\Omega_{;d}+\Omega \Omega_{;dc}\Bigr)
\gamma_{qr;pba}+\Bigr(\Omega_{;b}\Omega_{;d}
+\Omega \Omega_{;db}\Bigr)\gamma_{qr;pca} \cr
&+\Bigr(\Omega_{;a}\Omega_{;d}+\Omega \Omega_{;da}\Bigr)
\gamma_{qr;pcb}+\Bigr(\Omega_{;b}\Omega_{;c}
+\Omega \Omega_{;cb}\Bigr)\gamma_{qr;pda} \cr
&+\Bigr(\Omega_{;a}\Omega_{;c}+\Omega \Omega_{;ca}\Bigr)
\gamma_{qr;pdb}+\Bigr(\Omega_{;a}\Omega_{;b}
+\Omega \Omega_{;ba}\Bigr)\gamma_{qr;pdc}\biggr],
&(5.10)\cr}
$$
$$ \eqalignno{
\; & Z_{abcdpqr}^{(3)} \equiv 2 \biggr[\Bigr(\Omega_{;dc}\Omega_{;p}
+\Omega_{;d}\Omega_{;pc}+\Omega_{;c}\Omega_{;pd}
+\Omega \Omega_{;pdc}\Bigr)\gamma_{qr;ba} \cr
&+\Bigr(\Omega_{;db}\Omega_{;p}+\Omega_{;d}\Omega_{;pb}
+\Omega_{;b}\Omega_{;pd}+\Omega \Omega_{;pdb}\Bigr)
\gamma_{qr;ca} \cr
&+\Bigr(\Omega_{;da}\Omega_{;p}+\Omega_{;d}\Omega_{;pa}
+\Omega_{;a}\Omega_{;pd}+\Omega \Omega_{;pda}\Bigr)
\gamma_{qr;cb} \cr
&+\Bigr(\Omega_{;cb}\Omega_{;p}+\Omega_{;c}\Omega_{;pb}
+\Omega_{;b}\Omega_{;pc}+\Omega \Omega_{;pcb}\Bigr)
\gamma_{qr;da} \cr
&+\Bigr(\Omega_{;ca}\Omega_{;p}+\Omega_{;c}\Omega_{;pa}
+\Omega_{;a}\Omega_{;pc}+\Omega \Omega_{;pca}\Bigr)
\gamma_{qr;db} \cr
&+\Bigr(\Omega_{;ba}\Omega_{;p}+\Omega_{;b}\Omega_{;pa}
+\Omega_{;a}\Omega_{;pb}+\Omega \Omega_{;pba}\Bigr)
\gamma_{qr;dc} \cr
&+\Bigr(\Omega_{;cb}\Omega_{;d}+\Omega_{;c}\Omega_{;db}
+\Omega_{;b}\Omega_{;dc}+\Omega \Omega_{;dcb}\Bigr)
\gamma_{qr;pa} \cr
&+\Bigr(\Omega_{;ca}\Omega_{;d}+\Omega_{;c}\Omega_{;da}
+\Omega_{;a}\Omega_{;dc}+\Omega \Omega_{;dca}\Bigr)
\gamma_{qr;pb} \cr
&+\Bigr(\Omega_{;ba}\Omega_{;d}+\Omega_{;b}\Omega_{;da}
+\Omega_{;a}\Omega_{;db}+\Omega \Omega_{;dba}\Bigr)
\gamma_{qr;pc} \cr
&+\Bigr(\Omega_{;ba}\Omega_{;c}+\Omega_{;b}\Omega_{;ca}
+\Omega_{;a}\Omega_{;cb}+\Omega \Omega_{;cba}\Bigr)
\gamma_{qr;pd}\biggr],
&(5.11)\cr}
$$
$$ \eqalignno{
\; & Z_{abcdpqr}^{(4)} \equiv 2 \biggr[
\Bigr(\Omega_{;dcb}\Omega_{;p}+\Omega_{;dc}\Omega_{;pb}
+\Omega_{;db}\Omega_{;pc}+\Omega_{;d}\Omega_{;pcb} \cr
&+\Omega_{;cb}\Omega_{;pd} +\Omega_{;c}\Omega_{;pdb}
+\Omega_{;b}\Omega_{;pdc}+\Omega \Omega_{;pdcb}\Bigr)
\gamma_{qr;a} \cr
&+\Bigr(\Omega_{;dca}\Omega_{;p}+\Omega_{;dc}\Omega_{;pa}
+\Omega_{;da}\Omega_{;pc}+\Omega_{;d}\Omega_{;pca} \cr
&+\Omega_{;ca}\Omega_{;pd} +\Omega_{;c}\Omega_{;pda}
+\Omega_{;a}\Omega_{;pdc}+\Omega \Omega_{;pdca}\Bigr)
\gamma_{qr;b} \cr
&+\Bigr(\Omega_{;dba}\Omega_{;p}+\Omega_{;db}\Omega_{;pa}
+\Omega_{;da}\Omega_{;pb}+\Omega_{;d}\Omega_{;pba} \cr
&+\Omega_{;ba}\Omega_{;pd} +\Omega_{;b}\Omega_{;pda}
+\Omega_{;a}\Omega_{;pdb}+\Omega \Omega_{;pdba}\Bigr)
\gamma_{qr;c} \cr
&+\Bigr(\Omega_{;cba}\Omega_{;p}+\Omega_{;cb}\Omega_{;pa}
+\Omega_{;ca}\Omega_{;pb}+\Omega_{;c}\Omega_{;pba} \cr
&+\Omega_{;ba}\Omega_{;pc} +\Omega_{;b}\Omega_{;pca}
+\Omega_{;a}\Omega_{;pcb}+\Omega \Omega_{;pcba}\Bigr)
\gamma_{qr;d} \cr
&+\Bigr(\Omega_{;cba}\Omega_{;d}+\Omega_{;cb}\Omega_{;da}
+\Omega_{;ca}\Omega_{;db}+\Omega_{;c}\Omega_{;dba} \cr
&+\Omega_{;ba}\Omega_{;dc} +\Omega_{;b}\Omega_{;dca}
+\Omega_{;a}\Omega_{;dcb}+\Omega \Omega_{;dcba}\Bigr)
\gamma_{qr;p}\biggr],
&(5.12)\cr}
$$
$$ \eqalignno{
\; & Z_{abcdpqr}^{(5)} \equiv 2 \biggr[\Bigr(\Omega_{;dcba}\Omega_{;p}
+\Omega_{;cba}\Omega_{;pd}+\Omega_{;dba}\Omega_{;pc}
+\Omega_{;dca}\Omega_{;pb} \cr
&+\Omega_{;dcb}\Omega_{;pa}+\Omega_{;ba}\Omega_{;pdc}
+\Omega_{;ca}\Omega_{;pdb} \cr
&+\Omega_{;cb}\Omega_{;pda}+\Omega_{;da}\Omega_{;pcb}
+\Omega_{;db}\Omega_{;pca} \cr
&+\Omega_{;dc}\Omega_{;pba}+\Omega_{;a}\Omega_{;pdcb}
+\Omega_{;b}\Omega_{;pdca} \cr
&+\Omega_{;c}\Omega_{;pdba}+\Omega_{;d}\Omega_{;pcba}
+\Omega \Omega_{;pdcba}\Bigr)\gamma_{qr}\biggr].
&(5.13)\cr}
$$
The previous formulae imply that
$$ \eqalignno{
\; & S^{e}(\Omega^{2}\gamma) \equiv Q^{e}(\Omega^{2}\gamma)
+\nabla_{p}{\widetilde T}^{pe}(\Omega^{2}\gamma) \cr
&=\Omega^{2} Q^{e}(\gamma)+U^{e}
+\nabla_{p}{\widetilde T}^{pe}(\Omega^{2}\gamma).
&(5.14)\cr}
$$
On the other hand, our gauge is invariant under conformal rescalings
of $\gamma$ if and only if
$$
S^{e}(\Omega^{2}\gamma)=\Omega^{2} \Bigr(Q^{e}(\gamma)
+\nabla_{p}{\widetilde T}^{pe}(\gamma) \Bigr).
\eqno (5.15)
$$
By virtue of Eqs. (5.14) and (5.15), the desired tensor 
${\widetilde T}^{pe}(\gamma)$ should obey the equation
$$
\nabla_{p}{\widetilde T}^{pe}(\Omega^{2}\gamma)
-\Omega^{2}\nabla_{p}{\widetilde T}^{pe}(\gamma)
=-U^{e}(\gamma,\nabla^{(k)} \gamma, 
\Omega, \nabla^{(k)} \Omega).
\eqno (5.16)
$$
We now use the identity
$$
\nabla_{p}(\Omega^{2}{\widetilde T}^{pe}(\gamma))
=2\Omega \Omega_{;p}{\widetilde T}^{pe}(\gamma)
+\Omega^{2}\nabla_{p}{\widetilde T}^{pe}(\gamma)
\eqno (5.17)
$$
to re-express Eq. (5.16) in the form
$$
\nabla_{p} \Bigr({\widetilde T}^{pe}(\Omega^{2}\gamma)-\Omega^{2}
{\widetilde T}^{pe}(\gamma)\Bigr)
+2\Omega \Omega_{;p}{\widetilde T}^{pe}(\gamma)
=-U^{e}.
\eqno (5.18)
$$
At this stage we consider a vector $f^{e}$ on $(M,g)$, with 
associated covector $f_{e} \equiv g_{ep}f^{p}$ having 
non-vanishing contraction with $U^{e}$. The Leibniz rule
yields therefore
$$ \eqalignno{
\; & \nabla_{p}\Bigr(f_{e}{\widetilde T}^{pe}(\Omega^{2}\gamma)
-\Omega^{2}f_{e}{\widetilde T}^{pe}(\gamma)\Bigr)
-\Bigr({\widetilde T}^{pe}(\Omega^{2}\gamma)
-\Omega^{2}{\widetilde T}^{pe}(\gamma)
\Bigr)f_{e;p} \cr
&+2\Omega \Omega_{;p}f_{e}{\widetilde T}^{pe}(\gamma)
=-U^{e}f_{e}.
&(5.19)\cr}
$$
Thus, on defining
$$
C^{p} \equiv f_{e}{\widetilde T}^{pe}(\Omega^{2}\gamma)
-\Omega^{2}f_{e}{\widetilde T}^{pe}(\gamma),
\eqno (5.20)
$$
integration over $(M,g)$ of both sides of Eq. (5.19) and
application of the divergence theorem yield ($N_{b}$ being
the normal to $\partial M$)
$$ \eqalignno{
\; & \int_{\partial M} \Bigr[f_{e}{\widetilde T}^{pe}(\Omega^{2}\gamma)
-\Omega^{2}f_{e}{\widetilde T}^{pe}(\gamma)\Bigr]N_{p}d\sigma
=-\int_{M}U^{e}f_{e}dV \cr
&-2 \int_{M}\Omega \Omega_{;p}f_{e}{\widetilde T}^{pe}(\gamma)dV \cr
&+\int_{M}f_{e;p}\Bigr[{\widetilde T}^{pe}(\Omega^{2}\gamma)
-\Omega^{2}{\widetilde T}^{pe}(\gamma)\Bigr]dV,
&(5.21)\cr}
$$
where the left-hand side results from
$$
\int_{M}C_{\; \; ;p}^{p} \; dV.
$$
The tensor ${\widetilde T}^{pe}(\gamma)$ which solves the problem 
of finding a ``compensating term" 
$\nabla_{p}{\widetilde T}^{pe}(\gamma)$ such that the
conformal invariance condition (5.15) is fulfilled is therefore
obtained implicitly in non-local form, by solving the integral
equation (5.21). The integration measure $dV$ in Eq. (5.21) 
involves the background metric $g$ because we started from Eq.
(5.18) which contains covariant derivatives that annihilate $g$.
Note that, if $f^{e}$ were a Killing vector for $(M,g)$, only the
antisymmetric part of ${\widetilde T}^{pe}(\gamma)$ would 
contribute to the last integral in Eq. (5.21).

At this stage, the reader might still be wondering whether the 
proof of conformal invariance has been actually obtained,
because Eq. (5.21) is very complicated, and both members involve
integrals. We can now take advantage of the arbitrariness in
the choice of ${\widetilde T}^{pe}(\gamma)$, 
since it is sufficient to show that a class
of ${\widetilde T}^{pe}(\gamma)$ tensors exist for which Eq.
(5.18) (and hence Eq. (5.15)) is satisfied. For this purpose,
we {\it assume} that
$$
\nabla_{q}\Bigr[{\widetilde T}^{pe}(\Omega^{2}\gamma)-\Omega^{2}
{\widetilde T}^{pe}(\gamma)\Bigr]
+2\Omega \Omega_{;q}{\widetilde T}^{pe}(\gamma)
=-{1\over m}\delta_{q}^{\; p} U^{e}.
\eqno (5.22)
$$
Of course, Eq. (5.22) is not implied by Eq. (5.18), but on setting
$q=p$ and summing over $p$ one recovers Eq. (5.18). We can now
multiply both sides of Eq. (5.22) by $f^{q}$ and contract over
the index $q$. Upon setting
$$
{\cal D} \equiv f^{q}\nabla_{q}
\eqno (5.23)
$$
the resulting equation reads
$$
{\cal D}\Bigr[{\widetilde T}^{pe}(\Omega^{2}\gamma)-\Omega^{2}
{\widetilde T}^{pe}(\gamma)\Bigr]+2\Omega ({\cal D}\Omega)
{\widetilde T}^{pe}(\gamma)=-{1\over m}f^{p}U^{e}.
\eqno (5.24)
$$
This suggests applying the inverse operator ${\cal D}^{-1}$ to
both sides of Eq. (5.24) to solve for 
${\widetilde T}^{pe}(\gamma)$ with the
help of an integral equation. However, we have to turn the
resulting equation into an equation for differential forms,
since otherwise the integration over $M$ is ill-defined. For
this purpose, we define (the symbol $\otimes_{s}$ denotes
symmetrized tensor product)
$$
{\widetilde T}_{(\;)} \equiv {\widetilde T}_{pe}
dx^{p} \otimes_{s} dx^{e}={\widetilde T}_{(pe)}
dx^{p} \otimes_{s} dx^{e},
\eqno (5.25)
$$
$$
{\widetilde T}_{\wedge} \equiv {\widetilde T}_{pe}
dx^{p} \wedge dx^{e} = {\widetilde T}_{[pe]}
dx^{p} \wedge dx^{e},
\eqno (5.26)
$$
$$
(fU)_{(\;)} \equiv f_{p}U_{e}dx^{p} \otimes_{s} dx^{e}
=f_{(p} U_{e)} dx^{p} \otimes_{s}dx^{e},
\eqno (5.27)
$$
$$
(fU)_{\wedge} \equiv f_{p}U_{e} dx^{p} \wedge dx^{e}
=f_{[p} U_{e]} dx^{p} \wedge dx^{e}.
\eqno (5.28)
$$
Thus, since the operator ${\cal D}^{-1}$ is an integral operator
with kernel equal to the Green function $G_{{\cal D}}(x,y)$ of the
operator $\cal D$, we find
$$ \eqalignno{
\; & \Bigr[{\widetilde T}_{(\;)}(\Omega^{2}\gamma)
-\Omega^{2}{\widetilde T}_{(\;)}(\gamma)\Bigr](x)
+2 \int_{M}G_{\cal D}(x,y)[\Omega {\cal D} \Omega](y)
[{\widetilde T}_{(\;)}(\gamma)](y)dV(y) \cr
&=-{1\over m} \int_{M}G_{\cal D}(x,y)(fU)_{(\;)}(y)
dV(y),
&(5.29)\cr}
$$
$$ \eqalignno{
\; & \Bigr[{\widetilde T}_{\wedge}(\Omega^{2}\gamma)
-\Omega^{2}{\widetilde T}_{\wedge}(\gamma)\Bigr](x)
+2 \int_{M}G_{\cal D}(x,y)[\Omega {\cal D} \Omega](y)
[{\widetilde T}_{\wedge}(\gamma)](y)dV(y) \cr
&=-{1\over m} \int_{M}G_{\cal D}(x,y)
(fU)_{\wedge}(y) dV(y),
&(5.30)\cr}
$$
where $dV(y) \equiv \sqrt{{\rm det} g(y)} \; dy_{1}
... dy_{m}$ if $g$ is positive-definite. 
This form of the integral equations for the 
symmetric and anti-symmetric parts of ${\widetilde T}_{pe}$
suggests defining the kernel
$$
K_{\Omega}(x,y) \equiv 2 G_{\cal D}(x,y)[\Omega {\cal D}
\Omega](y)-\delta(x,y) \Omega^{2}(y),
\eqno (5.31)
$$
where $\delta(x,y)$ is the Dirac $\delta$-distribution. One can
therefore re-express Eqs. (5.29) and (5.30) in the form
$$ \eqalignno{
\; & \Bigr[{\widetilde T}_{\diamond}(\Omega^{2}\gamma)\Bigr](x)
+\int_{M}K_{\Omega}(x,y)\Bigr[{\widetilde T}_{\diamond}
(\gamma)\Bigr](y) dV(y) \cr
&=-{1\over m} \int_{M}G_{\cal D}(x,y)(fU)_{\diamond}(y)dV(y),
&(5.32)\cr}
$$
where the symbol $\diamond$ is a concise notation for the subscript
$(\;)$ used in Eq. (5.29) or the subscript $\wedge$ used in Eq.
(5.30). The right-hand side of Eq. (5.32) is completely known
for a given choice of the vector $f^{p}$ and of the dimension
$m$ of $M$ (see (5.7)--(5.13)).

We can now use a method similar to the one applied in the end
of Sec. 4. For this purpose, we assume that the metric $\gamma$
is positive-definite. If the operator $\cal D$ defined in (5.23)
is symmetric and elliptic on a compact Riemannian manifold without
boundary, it admits a discrete spectral resolution with
$C^{\infty}$ eigenvectors $\beta^{(n)}(x)$ satisfying the
eigenvalue equation
$$
{\cal D} \beta^{(n)}(x)=\lambda_{(n)} \beta^{(n)}(x).
\eqno (5.33)
$$
It is then possible to consider the expansions
$$
\Bigr[{\widetilde T}_{\diamond}(\gamma)\Bigr](x)
=\sum_{n=1}^{\infty}{\widetilde T}_{(n,\diamond)}(\gamma)
\beta^{(n)}(x),
\eqno (5.34)
$$
$$
\Bigr[{\widetilde T}_{\diamond}(\Omega^{2}\gamma)\Bigr](x)
=\sum_{n=1}^{\infty}{\widetilde T}_{(n,\diamond)}(\Omega^{2}\gamma)
\beta^{(n)}(x),
\eqno (5.35)
$$
where we assume that the right-hand sides have no part belonging
to the kernel of $\cal D$ (since otherwise the resulting algorithm
would not lead to algebraic equations, because $K_{\Omega}(x,y)$
does not annihilate any function $u$ such that ${\cal D}u=0$).
This is equivalent to choosing the vector $f^{e}$ so that the
resulting operator $\cal D$ (see (5.23)) has no zero-modes.
Moreover, on defining
$$
f(x) \equiv {1\over m} \int_{M}G_{\cal D}(x,y)
(fU)_{\diamond}(y)dV(y),
\eqno (5.36)
$$
$$
\rho^{(n)}(x) \equiv \int_{M}K_{\Omega}(x,y)
\beta^{(n)}(y)dV(y),
\eqno (5.37)
$$
we can further expand the coefficient $\rho^{(n)}(x)$ in terms
of the eigenvectors of $\cal D$ (cf. (4.11)), i.e.
$$
\rho^{(n)}(x)=\sum_{q=1}^{\infty}\rho^{(n)(q)}\beta^{(q)}(x),
\eqno (5.38)
$$
while for $f(x)$ we write
$$
f(x)=\sum_{n=1}^{\infty}f_{(n)}\beta^{(n)}(x).
\eqno (5.39)
$$
By virtue of (5.34)--(5.39), Eq. (5.32) is equivalent to the
infinite system of algebraic equations (cf. (4.13))
$$
\Bigr[{\widetilde T}_{(n,\diamond)}(\Omega^{2}\gamma)\Bigr]
+\sum_{q=1}^{\infty}{\widetilde T}_{(q,\diamond)}(\gamma)
\rho^{(q)(n)}=-f_{(n)}.
\eqno (5.40)
$$

In Eq. (4.3), the first integrand involves 
${\cal B}_{p}^{\; r}\varphi_{r}$. This is why the inclusion
of ${\overline \varphi}_{r}(x) \in {\rm Ker} \;
{\cal B}_{p}^{\; r}$ in the expansion (4.5) does not affect
the evaluation of the corresponding integral. However, in
Eq. (5.32) the first integrand does not involve any differential
operator acting on $\Bigr[{\widetilde T}_{\diamond}(\gamma)\Bigr](y)$.
This is why only the form (5.34) of the expansion leads to an
useful algorithm. One can then check that in the expansion (5.39)
any ${\overline f}(x) \in {\rm Ker} \; {\cal D}$ has to vanish, 
upon imposing Eq. (5.32) and the remaining expansions. In Eq.
(5.40) both $f_{(n)}$ and $\rho^{(q)(n)}$ are known. Hence one
gets an elegant but complicated solution, where the coefficients
${\widetilde T}_{(n,\diamond)}(\gamma)$ and
${\widetilde T}_{(n,\diamond)}(\Omega^{2}\gamma)$ obey an infinite
system of equations.
\vskip 0.3cm
\leftline {\bf 6. Concluding Remarks}
\vskip 0.3cm
\noindent
The original contributions of our paper are as follows.
\vskip 0.3cm
\noindent
(i) From the analysis of general mathematical structures (i.e.
vector bundles over Riemannian manifolds and conformal symmetries),
we have suggested that, if the Lorenz gauge can be replaced by the
Eastwood--Singer gauge in Maxwell theory, the de Donder gauge can 
be replaced by a broader family of supplementary conditions in
general relativity. 
\vskip 0.3cm
\noindent
(ii) A restriction on the parameter in the DeWitt supermetric (1.8)
has been obtained in a simple and elegant way. 
\vskip 0.3cm
\noindent
(iii) When the background is flat and Riemannian, i.e. 
$m$-dimensional Euclidean space ${\bf E}^{m}$, 
the admissibility of such gauges in
linearized theory has been proven by using the Green kernel of the 
Laplacian, the theory of polyharmonic functions and polyharmonic
forms, and Fourier transform techniques. The result (2.25) is 
non-local in that it involves integrals over ${\bf E}^{m}$. 
The solution $\varphi_{a}(x)$ is unique since the operator
$\cstok{\ }^{3}$ is elliptic.
\vskip 0.3cm
\noindent
(iv) When the background is flat and Lorentzian, i.e. 
$m$-dimensional Minkowski space-time, the admissibility of such gauges
in linearized theory has been proven by using the Green kernel
of the wave operator and partial Fourier transforms with respect to
time and space. The solution $\varphi_{a}({\vec x},t)$ depends on
the contour of integration, i.e. on suitable conditions imposed
on $\varphi_{a}({\vec x},t)$. We have considered, in particular,
a contour integral of the Feynman type, but it remains to be seen
whether other choices of contour integral give rise to solutions
which are more relevant for physics.
\vskip 0.3cm
\noindent
(v) When the background is curved with positive-definite metric,
the gauges (1.10) remain admissible provided that $\alpha \not
= -2$ (see (ii)) and if the operator ${\cal B}_{e}^{\; q}$ 
defined in (4.1) admits a discrete spectral resolution. 
Useful operational criteria to check the admissibility of our
gauges are then provided by (4.6) or (4.7) for a given choice
of Riemann curvature and of the tensor field $T^{pebc}$.
\vskip 0.3cm
\noindent
(vi) The condition (5.15) for conformal invariance of the gauge
conditions is fulfilled provided that 
${\widetilde T}^{pe}(\gamma)$ is chosen
to satisfy the integral equation (5.21). The counterpart of the
Eastwood--Singer gauge for gravity is therefore far more elaborated.
In particular, if the assumption expressed by Eq. (5.22) holds, we
have obtained the integral equation (5.32), and the solution is
found (in principle) by solving the infinite system of algebraic
equations (5.40).
\vskip 0.3cm
In our proof of conformal invariance of gauge conditions, it is
crucial to consider conformal rescalings of the physical metric
$\gamma_{ab}$, while the background metric $g_{ab}$ is kept
fixed. We have done so because it is $\gamma_{ab}$ which solves the
Einstein equations, which are not conformally invariant. The
consideration of general mathematical structures seems to suggest
that a key ingredient is the addition of a ``compensating term''
$\nabla_{p}{\widetilde T}^{pe}(\gamma)$ to the higher-order 
covariant derivatives of the original gauge condition (see
(1.5b) and (5.2)). Unlike the case of Maxwell theory in curved
backgrounds, where conformal rescalings of the background metric
are considered, we have therefore studied conformal rescalings
of the physical metric only in general relativity. Still, it remains
of interest for further research to consider conformal rescalings
of both background and physical metric.

Indeed, the importance of integral equations in general relativity
was already investigated, although from a completely different
perspective, in Ref. 13 (where the metric tensor was taken to be
linearly related to the energy-momentum tensor through an integral
involving a kernel), whereas equations similar to our Eq. (5.18)
occur when the formulation of conservation laws is considered in
Einstein's theory of gravity.$^{14}$ It now remains to be seen how
to extend the results of Secs. 4 and 5, proved for positive-definite
metrics, to the case of Lorentzian metrics, which are of course the
object of interest in general relativity. We can however point out
that the derivation of the integral equations (5.21) and (5.32) does
not depend on the signature of the metric, and hence the construction
of ${\widetilde T}^{pe}(\gamma)$ remains non-local also in the
Lorentzian case. The Green functions that one may want to use will
be distinguished by various boundary conditions (cf. Sec. 3),
and hence the Lorentzian framework will be actually richer 
in this respect.

The above results seem to suggest that new 
perspectives are in sight in the investigation of supplementary
conditions in general relativity. They might have applications both
in classical theory (linearized equations in gravitational wave 
theory, symmetry principles and their impact on gauge conditions),
and in the attempts to quantize the gravitational field (at least
as far as its semiclassical properties are concerned).
Hopefully, in the years to come it will become clear whether the
necessary mathematics can be used to rule out (or verify) the
existence of some properties of the universe (see, in particular,
the restrictions on the DeWitt supermetric found in Ref. 15, where
big-bang nucleosynthesis has been used as a probe of the geometry
of superspace).
\vskip 0.3cm
\leftline {\bf Acknowledgments}
\vskip 0.3cm
\noindent
This work has been partially supported by PRIN97 `Sintesi', and the
authors are indebted to Ivan Avramidi for enlightening correspondence.
\vskip 0.3cm
\leftline {\bf References}
\vskip 0.3cm
\item {1.}
R. Penrose and W. Rindler, {\it Spinors and Space-Time,
Vol. I: Two-Spinor Calculus and Relativistic Fields}
(Cambridge University Press, Cambridge, 1984).
\item {2.}
R. Penrose and W. Rindler, {\it Spinors and Space-Time,
Vol. II: Spinor and Twistor Methods in Space-Time Geometry}
(Cambridge University Press, Cambridge, 1986).
\item {3.}
R. S. Ward and R. O. Wells, {\it Twistor Geometry and
Field Theory} (Cambridge University Press, Cambridge, 1990).
\item {4.}
P. B. Gilkey, {\it Invariance Theory, the Heat Equation 
and the Atiyah--Singer Index Theorem} 
(Chemical Rubber Company, Boca Raton, 1995).
\item {5.}
M. J. Duff, {\it Class. Quantum Grav.} {\bf 11}, 1387 (1994).
\item {6.}
M. Eastwood and I. M. Singer, {\it Phys. Lett.} 
{\bf A107}, 73 (1985).
\item {7.}
G. Esposito, {\it Dirac Operators and Spectral Geometry},
Cambridge Lecture Notes in Physics, Vol. 12 
(Cambridge University Press, Cambridge, 1998).
\item {8.}
J. D. Jackson, {\it Classical Electrodynamics}
(John Wiley, New York, 1975).
\item {9.}
N. D. Birrell and P. C. W. Davies, {\it Quantum Fields in
Curved Space} (Cambridge University Press, Cambridge, 1982);
S. A. Fulling, {\it Aspects of Quantum Field Theory in
Curved Space-Time}, London Mathematical Society Student Texts,
Vol. 17 (Cambridge University Press, Cambridge, 1989).
\item {10.}
N. Aronszajn, {\it Polyharmonic Functions} 
(Clarendon Press, Oxford, 1984).
\item {11.}
G. Esposito, {\it Nuovo Cimento} {\bf B112}, 1405 (1997).
\item {12.}
S. W. Hawking and G. F. R. Ellis, {\it The Large-Scale Structure
of Space-Time} (Cambridge University Press, Cambridge, 1973).
\item {13.}
D. W. Sciama, P. C. Waylen and R. C. Gilman, {\it Phys. Rev.}
{\bf 187}, 1762 (1969).
\item {14.}
R. Penrose, {\it Proc. R. Soc. Lond.} {\bf A381}, 53 (1982).
\item {15.}
A. Llorente and J. P\'erez--Mercader, {\it Phys. Lett.}
{\bf B355}, 461 (1995).

\bye